\RequirePackage{fixltx2e}
\documentclass[aps,prb,reprint,floatfix, superscriptaddress]{revtex4-1}
\usepackage{latexsym}
\usepackage{amsmath}
\usepackage{amsfonts}
\usepackage{amssymb}
\usepackage{amsthm}
\usepackage{physics}
\usepackage{qcircuit}
\usepackage{graphicx}
\usepackage{dcolumn}
\usepackage{caption}
\usepackage{bm}
\usepackage{algorithm}
\usepackage{algpseudocode}
\usepackage{xcolor}

\newcommand{\eq}[1]{Eq.~(\ref{#1})} %
\def\be{\begin{equation}} %
\def\ee{\end{equation}} %
\def\bea{\begin{eqnarray}} %
\def\eea{\end{eqnarray}} %
\newcommand{\kh}{\hat \kappa}
\newcommand{\hx}{\hat x}
\newcommand{\hy}{\hat y}

\newcommand{\hn}{\hat n}
\newcommand{\hsz}{\hat S_z}
\newcommand{\HSS}{\hat S^2}
\newcommand{\LA}[1]{\mathfrak{#1}}
\begin{document}
\title{On the order problem in construction of unitary operators for the Variational Quantum Eigensolver} 

\author{Artur F. Izmaylov}
\email{artur.izmaylov@utoronto.ca}
\affiliation{Department of Physical and Environmental Sciences,
  University of Toronto Scarborough, Toronto, Ontario, M1C 1A4,
  Canada}
\affiliation{Chemical Physics Theory Group, Department of Chemistry,
  University of Toronto, Toronto, Ontario, M5S 3H6, Canada}
\author{Manuel D\'{i}az-Tinoco}
\affiliation{Department of Physical and Environmental Sciences,
  University of Toronto Scarborough, Toronto, Ontario, M1C 1A4,
  Canada}
\affiliation{Chemical Physics Theory Group, Department of Chemistry,
  University of Toronto, Toronto, Ontario, M5S 3H6, Canada}
\author{Robert A. Lang}
\affiliation{Department of Physical and Environmental Sciences,
  University of Toronto Scarborough, Toronto, Ontario, M1C 1A4,
  Canada}
\affiliation{Chemical Physics Theory Group, Department of Chemistry,
  University of Toronto, Toronto, Ontario, M5S 3H6, Canada}

\begin{abstract}
One of the main challenges in the Variational Quantum Eigensolver (VQE) framework is construction of the unitary 
transformation. The dimensionality of the space for unitary rotations of $N$ qubits is $4^N-1$, 
which makes the choice of a polynomial subset of generators exponentially difficult process. 
Moreover, due to non-commutativity of generators,
the order in which they are used strongly affects results. Choosing the optimal order in a particular subset of generators
requires testing the factorial number of combinations. We propose an approach based on the 
Lie algebra - Lie group connection and corresponding closure relations that systematically eliminates the order problem.  
\end{abstract}

\maketitle

\date{\today}

\section{Introduction}

One of the most promising applications of near-term universal gate quantum computation is solving the electronic 
structure problem. Currently, the most feasible route to this task is through the variational quantum eigensolver (VQE), \cite{Peruzzo2014} which is a hybrid technique involving an iterative minimization of the electronic expectation value 
\begin{align} \label{eq:E}
E = \min_{\boldsymbol{\tau}} \bra{\bar 0} \hat U^\dagger (\boldsymbol{\tau}) \hat H \hat U (\boldsymbol{\tau})\ket{\bar 0},
\end{align}
 involving quantum and classical computers. First, a classical computer suggests a trial unitary transformation 
 $\hat U (\boldsymbol{\tau})$ that is encoded on a quantum computer as a circuit operating on 
the initial state of $N$ qubits $\ket{\bar 0} \equiv \ket{0}^{\otimes N}$. 
 At the end of this circuit, one measures the expectation 
 value of the qubit-space Hamiltonian $\hat H$, which is iso-spectral to the electronic Hamiltonian
 in the second quantized form. The measured electronic energy is provided to the classical computer that generates 
 a next guess for the parameterized unitary transformation $\hat U (\boldsymbol{\tau})$. 
 These cycles serve to minimize the energy expectation value and to approach the true electronic ground state energy. 


Generators of $N$-qubit unitary operations up to a global phase correspond to 
$4^{N}-1$ basis vectors of the $\LA{su}(2^{N})$ Lie algebra, $N$-qubit Pauli products
\begin{align} \label{pauli_prod}
\hat P = \bigotimes_{j=1}^{N} \hat \sigma_j,
\end{align}
where $\hat \sigma_j$'s are the Pauli operators  $\{\hat x_j, \hat y_j, \hat z_j \}$ or the $2 \times 2$ identity $\hat 1_j$ for the $j^{\rm th}$
qubit. Thus, any element of the corresponding unitary group $SU(2^N)$ can be presented as
\bea\label{eq:U}
\hat U(\boldsymbol{\tau}) = \prod_{k=1}^{M} e^{i\tau_k \hat P_k},
\eea
where $M\le 4^N-1$. Generators $\hat P_k$ do not generally commute, therefore 
the individual exponents in \eq{eq:U} do not commute either. Yet, since all generators have been included, 
their order is not important. Any order will be able to represent any element of the 
$SU(2^N)$ Lie group. Only values of $\tau_k$'s are changing in representing a particular $SU(2^N)$ 
element with different orders in \eq{eq:U}. 

The representation in \eq{eq:U} is not a conventional representation of a Lie group, $\mathcal{G}$, as an exponential map of 
the corresponding Lie algebra, $g$, with generators $\{g_k\}$, 
\bea\label{eq:sE}
G = e^{\sum_k C_k g_k}, ~G \in \mathcal{G},
\eea
where $C_k$ are real or complex numbers.\cite{Gilmore_Exp} 
We can arrive to \eq{eq:U} considering the following additional properties 
of $SU(2^N)$ and the exponential map. $SU(2^N)$ is a compact Lie group and the corresponding 
algebra $\LA{su}(2^N)$ is compact as well. For compact Lie groups the exponential mapping of the Lie algebra 
is surjective.\cite{Knapp_expMsurj} Thus, using the conventional exponential mapping between $\LA{su}(2^N)$ and 
$SU(2^N)$ one can generate universal covering group which is simply connected (identical to $SU(2^N)$). 
For simply connected groups, one can use globally analytic Baker-Campbell-Hausdorff reparametrizations
of a single exponent (\eq{eq:sE}) to a disentangled form of multiple exponents as in \eq{eq:U}.\cite{Gilmore_BCH} 

In practice, one cannot use $M$ that scales exponentially with $N$, this would defeat the purpose of involving 
quantum computers for efficient representation of $\hat U$.\footnote {Note that quantum computers 
cannot perform $\exp(i\tau_k \hat P_k)$ as elementary circuit operations (gates) 
but there are compilers that present such operations as sequences of universal gates containing not 
more than polynomial in $N$ number of gates. Thus, we assume that polynomial in the number of $P_k$'s 
algorithms is a desirable goal.}
Thus, one needs efficient heuristics to make $\hat P_k$ selection so that $M$ scales polynomially with $N$.
Two main polynomial heuristics have been suggested for $\hat U$: 
1) methods based on the fermionic unitary forms restricted by the 
orbital excitation level and then transformed to the qubit space\cite{Peruzzo2014,Lee:2019/jctc/311,Sokolov2019,Evangelista:kz} (e.g. unitary coupled cluster singles and doubles (UCCSD)\cite{Peruzzo2014,Sokolov2019,Mizukami2019,Hempel2018})
and 2) qubit-space techniques, where $\hat P_k$'s are selected based on the energy gradients with respect to $\tau_k$'s.
\cite{Ryabinkin2018,Grimsley2019,qAdapt}
Methods in both categories provide particular subsets of $\hat P_k$'s in \eq{eq:U}
and thus lead to the order problem: different orders of exponents in 
\eq{eq:U} are not equivalent in their 
ability to lower the expectation values via optimizing corresponding $\tau_k$'s. Recently, several works have 
shown how large variations can be from changing the order.\cite{Grimsley:2019wz,Tranter:2019im,Evangelista:kz} 

In this work we provide an approach to remove the order problem by using algebraic closure of a $\hat P_k$ 
set into a Lie sub-algebra of $\LA{su}(2^N)$ by considering all possible commutators within the set. Considering 
Lie algebra - Lie group connection through the exponential map, 
the sub-algebra closure guarantees generation of a closed sub-group of $SU(2^N)$ 
where the order of exponents is not essential. Thus, even in the absence of commutativity, the closed 
algebraic structure to be introduced is sufficient for the order problem removing.   

\section{Theory}

\subsection{Lie algebraic consideration}

We assume that a set of $\hat P_k$'s is selected based on some energy lowering heuristic. Formally, this set  
$\mathcal{S} = \{\hat P_k\}_{k=1}^M$ forms a Lie sub-algebra of $M$ generators if for any pair $\hat P_i$ and $\hat P_j$ 
\bea
[\hat P_i,\hat P_j] = \sum_k c_{ij}^{(k)} \hat P_k, ~\hat P_i,\hat P_j, \hat P_k \in \mathcal{S}
\eea
where $c_{ij}^{(k)} $ are so-called structural constants.\cite{Gilmore:2008}  
Then the exponential map of $\mathcal{S}$ produces a Lie sub-group $\mathcal{G}_S$ with elements
that can be given by \eq{eq:U}. The order of $\hat P_k$'s forming the $\mathcal{S}$ sub-algebra does not matter because 
any order of their exponential products can represent points of a sub-manifold corresponding to $\mathcal{G}_S$.
Here, again, one of the crucial elements is compactness of both $\mathcal{S}$ and $\mathcal{G}_S$.

If a single sub-algebra will not be sufficient to provide enough elements to lower the energy, one can use a set of sub-algebras,
$\{\mathcal{S}_j\}$ to construct $\hat U = \hat U_1 \times ... \times \hat U_n$, where  
 \bea \label{eq:Uj}
\hat U_j(\boldsymbol{\tau}_j) = \prod_{k=1}^{M_j} e^{i\tau_k^{(j)} \hat P_k^{(j)}},~ \hat P_k^{(j)} \in \mathcal{S}_j.
\eea
In this case, the order of exponents within each $\hat U_j$ will not affect the minimum of the energy expectation value,
but the order of $\hat U_j$'s in $\hat U$ will. This provides a simple prescription of how to remove the order 
problem for any neighbouring pairs $\hat U_j$ and $\hat U_{j+1}$: one needs to obtain the closure of corresponding 
sub-algebras $\mathcal{S}_j$ and $\mathcal{S}_{j+1}$, $\mathcal{S}_{j,j+1}^{(c)} = [\mathcal{S}_j,\mathcal{S}_{j+1}]$, 
which is a sub-algebra that contains all elements of $\mathcal{S}_j$ and $\mathcal{S}_{j+1}$ 
as well as all possible commutators. Clearly, the number of generators of 
$\mathcal{S}_{j,j+1}^{(c)}$ will not be smaller than or equal to the sum of those for 
$\mathcal{S}_j$ and $\mathcal{S}_{j+1}$. $\mathcal{S}_{j,j+1}^{(c)}$ elements can be exponentiated 
to create $\hat U_{j,j+1}$, which will be an order invariant substitute for $\hat U_j \hat U_{j+1}$.

\subsection{Efficiency consideration}

There are two questions that appear from the described closure procedure. 
First, how to choose sub-algebras so that the energy will be reduced efficiently? Second,
is there a way to reduce the growth of the number of terms during generation of various closures? 

To address the first question we can consider several heuristics for finding $\hat U$, they generally introduce 
the order dependence problem but with generating algebraic closures this dependency can be removed.
The main ideas behind these heuristics are: 1) fermionic excitations (one-electron or mean-field picture), and 
2) gradients along the $\hat P_k$'s (basis vectors in tangential space for the $SU(2^N)$ manifold). 

For the second question, starting with a sub-algebra whose size is desirable to reduce, one can search 
for linear combinations of $P_k$'s so that they commute with known symmetry operators. 
It is straightforward to show that if the Hamiltonian of the system has symmetry operators, $\{\hat S_k\}$ 
then a set of $\hat P_j$'s commuting with the symmetry operators form a sub-algebra, $\mathcal{S}_S$. 
It is enough to show that a commutator of any two elements of $\mathcal{S}_S$ also commutes with 
the symmetry operator and thus is in $\mathcal{S}_S$. To show this commutativity, 
we invoke the Jacobi condition for the commutator operation    
\bea
[\hat S_k, [\hat P_i,\hat P_j]] + [\hat P_i, [\hat P_j,\hat S_k]] + [\hat P_j, [\hat S_k,\hat P_i]] = 0.
\eea
Since both $\hat P_i$ and $\hat  P_j$ commute with $\hat S_k$ their commutator commutes with 
$\hat S_k$ as well. This consideration extends to any linear combination of $\hat  P_k$'s since 
the commutator is a bilinear operation. Using the elements of $\mathcal{S}_S$ in the exponential map one can generate
symmetry-adapted rotations that form a Lie sub-group. The number of generators in 
$\mathcal{S}_S$ cannot be larger than that in the original sub-algebra. Therefore, using 
symmetries like the number of electrons $\hat N_e$, the total electronic spin $\hat S^2$, and 
its $z$-projection $\hat S_z$, one can reduce the size of sub-algebras and 
construct sub-groups that will preserve the symmetries in unitary rotations.   

Another approach to the Lie algebra reduction is through its formal decompositions (e.g. Levi decomposition).\cite{Barut_p17} 
Since we work with sub-algebras of the compact Lie algebra $\LA{su}(2^N)$, one can use the following decomposition 
theorem: {\it any compact Lie algebra is a direct sum of the algebra center and simple sub-algebras.}\cite{Barut_p17}    
The algebra center commutes with all elements of the algebra, hence, based on Schur's
 lemma, action of the center elements on the wavefunction is equivalent to multiplication by a 
 constant if the wavefunction is within a single irreducible representation. Since we exponentiate the algebra elements 
 to obtain unitary transformations, one can remove the center elements from the construction if the wavefunction 
 has all components within one irreducible representation of the center elements.    

\subsection{Two types of generators}

A notion of a generator has some ambiguity when it is applied to Lie algebras. On the one hand, Lie algebras 
are linear spaces with $\hat P_k$'s as the basis elements (vectors). One can generate Lie groups 
via exponentiation of $\hat P_k$'s, and thus, it is reasonable to refer to all $\hat P_k$'s as generators of the associated Lie group. 
On the other hand, Lie algebras have a vector-vector multiplication operation, which is the commutator of two 
$\hat P_k$'s. Therefore, one can generate algebra basis vectors from a smaller subset of $\hat P_k$'s, which can be seen
as the algebra generators. Both types of generators can be useful, therefore, to distinguish 
them we will refer to the latter generators 
as generators of the second type (generators II). 
 To illustrate the difference between the two types of generators, let us consider a single-qubit case: there are three 
generators of the $SU(2)$ group, $\hat x$, $\hat y$, and $\hat z$, so that any single-qubit rotation can be written as 
\bea
\hat U_1 =  e^{i\theta_x \hat x}e^{i\theta_y \hat y}e^{i\theta_z \hat z}
\eea
and two generators II, any two operators out of the triplet ($\hat x$, $\hat y$, $\hat z$) can generate the 
third operator via commutation. The generators II give rise to the Euler angle parametrization of 
single-qubit rotations
\bea
\hat U_1' =  e^{i\theta_z' \hat z}e^{i\theta_y \hat y}e^{i\theta_z \hat z},
\eea 
note that repetition of some generators II is needed for a complete description of all rotations. This repetition 
can be considered as a compensation for a smaller set of exponentiated operators. 
It was shown previously, that the number 
of generators II scales only quadratically with $N$.\cite{Ryabinkin2018} Essentially, to construct all $4^N-1$ 
 generators of $\LA{su}(2^N)$ one needs only a subset of one- and two-qubit $P_k$'s, for example, 
 $\hat x$ and $\hat z$ operators for each qubit and pair of qubits. Similar considerations for the algebra 
of fermionic excitations shown that single and double excitations are sufficient to generate all possible 
excitations.\cite{Evangelista:kz} Note that due to necessity for repeated exponentiation of generators II, 
they create the most severe order problem in representation of unitaries, and we will not consider them further. 
%

\subsection{Unitary Coupled Cluster (UCC) Sub-algebras}

Historically, the UCC method was formulated using an exponent of a sum
for the unitary transformation,
\bea\label{eq:UCC}
\hat U = \exp[ \sum_n \hat T_n - \hat T_n^\dagger].
\eea 
Here, $\hat T_n$ are $n$-tuple excitations
\bea\label{eq:Tn}
\hat T_n = \sum_{ij...k,ab...c} t_{ij...k}^{ab...c} a_a^\dagger a_b^\dagger ... a_c^\dagger a_i a_j ... a_k,
\eea
where we assume $i,j,...k$ ($a,b,...c$) indices to correspond to occupied (unoccupied) molecular orbitals obtained  
in the Hartree-Fock method.
The number of amplitudes $t_{ij...k}^{ab...c}$ grows combinatorially with the number of orbitals and electrons, 
therefore to have polynomial heuristics $n$ is usually restricted to a fixed value, for example, $n=1$ (singles, S) 
\bea
 \hat T_1 = \sum_{ia} t_{i}^{a} a_a^\dagger a_i.
\eea
and $2$ (doubles, D) 
\bea
 \hat T_2 = \sum_{ijab} t_{i,j}^{a,b} a_a^\dagger a_b^\dagger a_i a_j.
\eea
This gives rise to the UCCSD parametrization,
\bea\label{eq:UCCSD}
\hat U_{\rm SD} = \exp[ \sum_{n=1}^2 \hat T_n - \hat T_n^\dagger].
\eea
All $\hat T_n$'s in \eq{eq:Tn} commute with each other component-wise, 
only addition of $\hat T_n^\dagger$ breaks down the commutativity. 
This poses a problem of presenting 
Eqs.~(\ref{eq:UCCSD}) and (\ref{eq:UCC}) as a product of exponents of individual excitation/de-excitation pairs.
The latter is needed for transferring the unitary transformation to a sequence of gates by compilers.    
Therefore, to implement UCCSD, one needs to do a Trotter approximation 
\bea\label{eq:SDTr}
\hat U_{\rm SD} \approx \left[\prod_{ia} e^{t_i^a \kh_i^a/K} \prod_{ijab} e^{t_{ij}^{ab} \kh_{ij}^{ab}/K} \right]^K,
\eea
where $K$ is a finite number of Trotter steps, and  
\bea
\kh_i^a &=& \hat E_{i}^a - \hat E_{a}^{i} = a_a^\dagger a_i -a_i^\dagger a_a\\
\kh_{ji}^{ab} &=& \hat E_{ji}^{ab} - \hat E_{ab}^{ji} =  a_a^\dagger a_b^\dagger a_i a_j - a_j^\dagger a_i^\dagger a_b a_a
\eea
are anti-hermitized single and double excitations. Note that $\kh$ do not commute, and order of 
terms in \eq{eq:SDTr} matters for any finite $K$. 

Using commutativity of $\hat T_n$ components one can ``unitarize" the coupled cluster excitation operator 
\bea
\hat \Omega_{\rm SD} &=& \exp\left[ \hat T_1 + \hat T_2 \right] \\
&=& \left[\prod_{ia} e^{t_i^a \hat E_i^a} \prod_{ijab} e^{t_{ij}^{ab} \hat E_{ij}^{ab}} \right]. 
\eea
as a product of exponents 
\bea
\hat U_{\rm dSD} &=& \left[\prod_{ia} e^{t_i^a \kh_i^a} \prod_{ijab} e^{t_{ij}^{ab} \kh_{ij}^{ab}} \right],
\eea
which is called disentangled UCCSD (dUCCSD) in Ref.~\citenum{Evangelista:kz}.
dUCCSD will not require the Trotter approximation but its results will depend on the order of terms as was 
demonstrated numerically in Ref.~\citenum{Evangelista:kz}.  

To address the order problem in the UCC case we consider possible sub-algebras originating 
from sets of fermionic excitations/de-excitations $\kh$.   

\paragraph{Individual unitary fermionic excitations/de-excitations:}  
Here, we study sets of Pauli products obtained from transforming 
individual single and double fermionic excitations/de-excitations 
using the Jordan-Wigner (JW) transformation. This transformation was chosen for
concreteness, similar results can be obtained using the parity or Bravyi-Kitaev transformations.  
It has been known that the individual fermionic excitation/de-excitation pairs produce sets of commuting 
Pauli products.\cite{Romero2018} In case of singles and doubles these are
\bea
&& \kh_{i}^{a} =  \frac{i}{2} \bigotimes_{k=i+1}^{a-1} \hat{z}_k (\hy_i\hx_a-\hx_i\hy_a)\\
&&\kh_{ij}^{ba} = \frac{i}{8} 
\bigotimes_{k=i+1}^{j-1} \hat{z}_k \bigotimes_{l=a+1}^{b-1} \hat{z}_l  
(\hat{x}_i \hat{x}_j \hat{y}_a \hat{x}_b +  \hat{y}_i \hat{x}_j \hat{y}_a \hat{y}_b\nonumber \\
 &&  + \hat{x}_i \hat{y}_j \hat{y}_a \hat{y}_b 
  + \hat{x}_i \hat{x}_j \hat{x}_a \hat{y}_b  - \hat{y}_i \hat{x}_j \hat{x}_a \hat{x}_b - \hat{x}_i \hat{y}_j \hat{x}_a \hat{x}_b\nonumber \\
&& - \hat{y}_i \hat{y}_j \hat{y}_a \hat{x}_b   - \hat{y}_i \hat{y}_j \hat{x}_a \hat{y}_b ). 
\eea
One can continue this for higher excitations/de-excitations, where commutativity of Pauli terms will persist. 
The origin of this commutativity is in the even number of differences ($\hat x$ / $\hat y$ operators) in Pauli 
products surviving the subtraction of the de-excitation part from the excitation part (see more details in 
the Appendix of Ref.~\citenum{Romero2018}). All commuting Pauli products within an excitation/de-excitation set 
form an abelian sub-algebra, which obviously removes the order problem within this set.

All excitations/de-excitations commute with the number of electrons operator, $\hat N_e$, because they 
do not change the number of electrons (the number of $a$ and $a^\dagger$ operators is the same). 
If one considers only $\hat N_e$-symmetry, the number of Pauli products within double excitations/de-excitations 
can be reduced to four terms which are mutually commutative
\bea
&&\hat \xi_{ij}^{ba}  = \frac{i}{4} 
\bigotimes_{k=i+1}^{j-1} \hat{z}_k \bigotimes_{l=a+1}^{b-1} \hat{z}_l  \notag \\ 
&&(\hat{x}_i \hat{x}_j \hat{y}_a \hat{x}_b +  \hat{y}_i \hat{x}_j \hat{y}_a \hat{y}_b
 - \hat{x}_i \hat{y}_j \hat{x}_a \hat{x}_b  - \hat{y}_i \hat{y}_j \hat{x}_a \hat{y}_b ), \\
&&\hat \pi_{ij}^{ba} = \frac{i}{4} 
\bigotimes_{k=i+1}^{j-1} \hat{z}_k \bigotimes_{l=a+1}^{b-1} \hat{z}_l  \notag  \\
&&(\hat{x}_i \hat{y}_j \hat{y}_a \hat{y}_b + \hat{x}_i \hat{x}_j \hat{x}_a \hat{y}_b 
- \hat{y}_i \hat{x}_j \hat{x}_a \hat{x}_b - \hat{y}_i \hat{y}_j \hat{y}_a \hat{x}_b ).
\eea
Single excitations are not reducible without violating commutativity with $\hat N_e$, higher excitations 
($n$-tuple with $n>1$) are reducible if only commutativity with $\hat N_e$ is required. 

Operators that commute with the electron spin operators, $\hsz$ and $\HSS$, can be obtained by 
anti-hermitization of singlet spherical tensor operators. General spherical tensor operator $\hat T^{S,M}$
is defined as 
\bea
[\hat S_{\pm}, \hat T^{S,M}] = \sqrt{S(S+1)-M(M\pm 1)} \hat T^{S,{M\pm 1}},
\eea
\bea
[ \hsz, \hat T^{S,M} ] = M \hat T^{S,M} ,
\eea 
where $S$ and $M$ are electron spin and its projection to the $z$-axis, respectively. 
Equation $\HSS = \hat S_{-}\hat S_{+} + \hsz(\hsz+1)$ can be used to show that any 
singlet spherical tensor operator, $\hat T^{0,0}$ will commute with $\hsz$ and $\HSS$. 

There are standard approaches for producing spherical tensor operators,\cite{Helgaker}
they involve very similar techniques to those used for generating spin-adapted 
configuration state functions.\cite{Paldus,Shavitt}
Individual single excitations are not $\hat T^{0,0}$ operators, therefore, one needs to group 
more than one excitation to obtain the singlet operator
\bea
\hat T^{0,0}_{ia,\bar{i}\bar{a}} = \kh_{i}^{a} +\kh_{\bar{i}}^{\bar{a}},
\eea
here and further $a(\bar{a})$ and $i(\bar{i})$ are spin-orbitals that have $\alpha$($\beta$) spin parts.
For double and higher excitations/de-excitations, one can use seniority (the number of unpaired electrons
created by the operator) to assess whether an individual excitation/de-excitation pair 
can be a singlet operator. Zero seniority always provides a positive answer, for example, zero-seniority 
double excitation/de-excitation gives 
\bea
\hat T^{0,0}_{i\bar{i}a\bar{a}} = \kh_{i\bar{i}}^{a\bar{a}}.
\eea
For seniority two, one needs to combine two double excitations/de-excitations
to obtain the singlet operator
\bea
\hat T^{0,0}_{i\bar{i}a\bar{b},i\bar{i}b\bar{a}} = \kh_{i\bar{i}}^{a\bar{b}} + \kh_{i\bar{i}}^{b\bar{a}}.
\eea 
Generally, the seniority number is equal to the number of individual excitation/de-excitation pairs in construction of 
singlet operators. It is worth noting that in all odd excitations, due to nonzero seniority, individual 
fermionic excitations/de-excitations do not form proper spin-conserving operators. This illustrates that imposing 
symmetry constraints to some sub-algebras can give empty sets and remove these sub-algebras from 
consideration too prematurely. 
   

\paragraph{All n-tuple excitations/de-excitations:} 
Among of all sets with a fixed rank of excitation only singles form closed sub-algebras
\bea
[\hat E_r^s,\hat E_p^q] = \hat E_r^q \delta_{ps} - \hat E_p^s \delta_{rq},
\eea
where we use $p,q,r,s,...$ indices for both occupied and unnocupied orbitals. 
This equation can be generalized for $\kh_r^s$'s. 
Starting from double excitations, commutators produce higher excitations, and therefore, fixed-excitation algebras
do not exist 
\bea
[\hat E_{pq}^{rs},\hat E_{tu}^{vw}] = \delta_{pw} \hat E_{tuq}^{rsv} - \delta_{pv} \hat E_{tuq}^{rsw} + ...
\eea
where $E_{tuq}^{rsv}=a_r^\dagger a_s^\dagger a_v^\dagger a_q a_u a_t$, and $...$ contain other triple excitations
and all double and single excitations. This consideration gives the same conclusion 
for anti-hermitized excitation/de-excitation pairs, hence, results of any finite order disentangled UCC beyond singles 
(e.g. dUCCSDT) will be order dependent.  Interestingly, even if one adds rotations within occupied and unoccupied
subspaces, a finite operator rank is not enough to provide a closure, therefore,
works using finite rank generalized UCC need Trotter steps as well.\cite{Lee:2019/jctc/311} 

\paragraph{Orbital restricted subsets of excitations/de-excitations:}

One way to obtain closed algebras is to restrict the number of orbitals involved 
in possible excitations and de-excitations. If no restriction is made then operators involving 
all orbitals will be needed to close the algebra. Generally, one would not only need 
operators involving excitations from occupied to unoccupied orbitals and corresponding de-excitations 
but also rotations within occupied and unoccupied spaces. 

\subsection{Qubit Coupled Cluster (QCC) Sub-algebras}

Another approach to construction of the unitary transformation is through a direct selection of Pauli products skipping 
the fermionic picture. Indeed, search for an element of the $SU(2^N)$ group minimizing the energy function [\eq{eq:E}] 
can be done using energy gradients with respect to individual amplitudes, $\tau_k$, in the representation of \eq{eq:U} 
\bea\label{eq:g}
\frac{\partial E}{\partial \tau_k} \Big|_{\boldsymbol{\tau} = 0} = i\bra{\bar{0}} [\hat H, \hat P_k] \ket{\bar{0}}, \label{energy_gradient_expr}
\eea
In the QCC method\cite{Ryabinkin2018} 
these gradients are evaluated using the qubit mean-field reference state, for simplicity we will take the 
Hartree-Fock Slater determinant as the reference state, $\ket{\bar{0}}$. In the qubit space, $\ket{\bar{0}}$ corresponds 
to a collinear product of qubits aligned with the $z$-axis. 
Such a reference state simplifies the selection of $\hat P_k$'s based on the energy gradient.\cite{Ryabinkin2019b} 

\paragraph{Equivalence classes based on the energy gradients:} 
We refer to the set of all $\hat P_k$'s with nonzero absolute energy gradients in \eq{eq:g} 
as the \textit{direct interaction set} (DIS), denoted herein as $\mathcal{D}$. 
The DIS is formed by $n_p$ nonoverlapping subsets (equivalence classes), $\mathcal{D} = \bigcup_{k=1}^{n_p} \mathcal{G}_k$, 
where each subset $\mathcal{G}_k$ contains $O(2^{N-1})$ $\hat P_k$'s of identical gradient magnitude.\cite{Ryabinkin2019b} 
The number of classes $n_p$ for the qubit Hamiltonian scales linearly in the number of terms it contains in the 
fermionic representation, and hence efficient ranking of all $\hat P_k$'s in the DIS is accomplished by performing a 
gradient calculation only for a single representative $\hat P_k$ for each class, thereby ranking the complete DIS of cardinality 
$O(2^{N-1}n_p)$ with merely $n_p$ gradient computations.

The nonzero gradient condition selects only $\hat  P_k$'s that contain not higher than fermionic 2-body interactions
due to 2-body terms in the Hamiltonian. $\hat P_k'$s from the same gradient class commute and thus form a sub-algebra. 
The sub-algebra of each nonzero gradient class is a linear subspace from which double fermionic excitation/de-excitation pairs 
can be built. Adding an extra condition of commutation with symmetry operators $\hat N_e$, $\hat S^2$, and $\hat S_z$
provides symmetry adapted combinations identical to those obtained in the double fermionic excitation/de-excitation case.  

To go beyond fermionic doubles in this scheme one can generate new DIS by either modifying the Hamiltonian 
or the reference state using unitary transformations built from high-gradient class $\hat P_k$'s. 
These approaches were used in recently suggested 
computational techniques, iterative QCC\cite{Ryabinkin2019b} and ADAPT-VQE.\cite{Grimsley2019,qAdapt} 

\paragraph{Sub-algebras based on anti-commuting sub-sets:}
To create sub-algebras that are very different from those obtained in the fermionic excitation/de-excitation case one needs to 
abandon symmetry considerations for a moment. 
Using DIS sets, it is possible to construct closed Lie sub-algebras starting with mutually anti-commutative Pauli products.  
Selecting $P_k$'s from different gradient classes, one can form a mutually anti-commutative set $\{\hat P_i\}$, where 
$\hat P_i \hat P_j = -\hat P_j \hat P_i$. A finite sub-algebra originates from these anti-commuting terms if 
all product pairs $\hat P_i\hat P_j$ are added to the set. Due to the involutory property of 
$\hat P_k$'s, $\hat P_k^2 = \hat 1$, all anti-commuting terms and their product pairs form a closed Lie sub-algebra
\begin{align}
[\hat P_i, \hat P_j] &= 2\hat P_i \hat P_j \\
[\hat P_i\hat P_j,\hat P_k] &= 2(\delta_{jk}\hat P_i-\delta_{ik}\hat P_j), \\
[\hat P_i\hat P_j,\hat P_k \hat P_l] &= 2(\delta_{jk} \hat P_i\hat P_l - \delta_{ik} \hat P_j\hat P_l \notag \\
&+\delta_{il} \hat P_j\hat P_k  -\delta_{jl} \hat P_i\hat P_k), 
\end{align}
The maximal size of these sub-algebras scales quadratically with the number of qubits because 
the maximal number of fully anti-commuting terms scales linearly with the number of qubits.\cite{Zhao2019}
There are two connections of this Lie algebra with other known algebras\cite{PaldusSarma1985}: 
1) $\{\hat P_i\}_{i=1}^K$ generate associative 
Clifford algebras, $C_K$, with ${\rm dim} (C_K) = 2^K$; 2) defining $\hat S_{0j}\equiv -\hat S_{j0}\equiv -i\hat P_j/2$
and $\hat S_{jk} \equiv [\hat P_j, \hat P_k]/4= \hat P_j\hat P_k /2$, where $j\ne k=1,...,K$ provides the special 
orthogonal Lie algebra $\LA{so}(K+1)$, which is compact and has the commutation relations
\bea
[\hat S_{ij},\hat S_{kl}] &=& \delta_{jk}\hat S_{il} + \delta_{il} \hat S_{jk} - \delta_{ik} \hat S_{jl} - \delta_{jl}\hat S_{ik}.
\eea

Interestingly, all sub-algebras constructed based on anti-commutative terms do not have finite symmetry adapted
sub-algebras. Any attempt to find symmetrized linear combinations with respect to electron spin and number operators 
lead to empty sets. To obtain symmetrized operators one needs to combine multiple anti-commutativity-based sub-algebras into 
a composite algebra, this composite algebra will usually contain symmetry adapted sub-algebras. However, the symmetry adapted 
sub-algebras obtained this way are not different from those generated using symmetrization of fermionic excitations/de-excitations. 

Another structural feature of the Pauli products algebra is that any sub-set of commuting operators naturally constitutes 
a Lie sub-algebra, and any fully anti-commuting subset can be organized into a Lie sub-algebra if all products are added, 
but if a subset contains commuting and anti-commuting elements then its closure may require an exponential number of elements. 
A simple illustration of the last case is that a set of all one- and two-qubit operators for 
$N$-qubits can generate all $4^N-1$ elements of the $\LA{su}(2^N)$ algebra. 
    


\section{Model example}  \label{sec:benchmarking}

To illustrate two main steps in eliminating the order problem we consider the 
recently proposed model example of two electrons 
within a space of four spin-orbitals, two occupied $i,\bar{i}$ 
and two unoccupied $a,\bar{a}$.\cite{Evangelista:kz}  
Thus, an arbitrary state of two electrons can be written as
\bea\label{eq:FCI}
| \Psi \rangle = c_1 | i \bar{i} \rangle + c_2 | a \bar{i} \rangle + c_3 | \bar{a} i \rangle + c_4 | a\bar{a} \rangle
\eea
where $\ket{i \bar{i}}, \ket{a \bar{i}}, \ket{\bar{a} i}, \ket{a\bar{a}}$ are four Slater determinants and $c_i$'s are their 
normalized coefficients, $\sum_i |c_i|^2=1$. 
It is possible to construct  three fermionic excitation/de-excitation operators: $\kh_{i\bar{i}}^{a\bar{a}}, \kh_{\bar{i}}^{\bar{a}},\kh_{i}^{a}$.
Interestingly, the order in which these three operators are placed in the 
disentangled UCC wavefunction can strongly affect the result. These differences are so dramatic that one 
particular order when applied to the physical vacuum state $\ket{i \bar{i}}$ 
\begin{align}\label{eq:P211}
\ket{\Psi_{211}} =  e^{t_3 \kh_{i\bar{i}}^{a\bar{a}}} e^{t_2 \kh_{\bar{i}}^{\bar{a}}} e^{t_1 \kh_{i}^{a}} \ket{i\bar{i}}
\end{align}
cannot represent an arbitrary state by varying the amplitudes $t_i$, while different orders like the following 
\begin{align}
\ket{\Psi_{121}} = e^{t_2\hat{\kappa}_{\bar{i}}^{\bar{a}}} e^{t_3 \hat{\kappa}_{i\bar{i}}^{a\bar{a}}} e^{t_1 \hat{\kappa}_{i}^{a}} |i\bar{i} \rangle
\end{align}
can. This order dependence is easy to understand if one creates the algebraic closure with respect to commutation
for the set of the three fermionic operators. This closure results in the Lie algebra containing 8 operators: 
\bea
\mathcal{A} &=& \{\kh_{i\bar{i}}^{a\bar{a}}, \kh_{\bar{i}}^{\bar{a}}, \kh_{i}^{a}, \kh_{i\bar{a}}^{a\bar{i}}, 
(\hn_a-\hn_i)\kh_{\bar{i}}^{\bar{a}},(\hn_{\bar{a}}-\hn_{\bar{i}})\kh_{i}^{a},\notag\\
&& (\hn_a-\hn_i)^2\kh_{\bar{i}}^{\bar{a}},(\hn_{\bar{a}}-\hn_{\bar{i}})^2\kh_{i}^{a}\},
\eea
where $\hn_p = a_p^\dagger a_p$. This algebra can be decomposed as a direct sum of the center 
\bea
\mathcal{C} = \{ [1-(\hn_a-\hn_i)^2]\kh_{\bar{i}}^{\bar{a}},[1-(\hn_{\bar{a}}-\hn_{\bar{i}})^2]\kh_{i}^{a} \}
\eea
and two simple algebras isomorphic to $\LA{su}(2)$
\bea
 \mathcal{A}_1 &=& \{ (\hn_{\bar{a}}-\hn_{\bar{i}})\kh_{i}^{a}, (\kh_{i\bar{i}}^{a\bar{a}}+\kh_{i\bar{a}}^{a\bar{i}}), (\hn_a-\hn_i)^2\kh_{\bar{i}}^{\bar{a}}\} \\
 \mathcal{A}_2 &=& \{(\hn_a-\hn_i)\kh_{\bar{i}}^{\bar{a}}, (\kh_{i\bar{i}}^{a\bar{a}}-\kh_{i\bar{a}}^{a\bar{i}}), (\hn_{\bar{a}}-\hn_{\bar{i}})^2\kh_{i}^{a}\}.
\eea
All elements of $\mathcal{C}$ zero any configuration with two electrons, hence, to avoid the order problem 
one needs to use only elements from $\mathcal{A}_1$ and $\mathcal{A}_2$. 

Further reduction of the number of operators can be achieved by imposing the symmetry constraints to obtain 
symmetry adapted sub-algebras. 
All elements of $\mathcal{A}$ commute with $\hat N_e$ and $\hat S_z$, but to introduce the commutativity 
with $\hat S^2$ one needs to combine a few elements of $\mathcal{A}$ together. 
The symmetry-adapted set of operators forming a Lie sub-algebra $\{\hat A_i\}$ is 
\begin{align}
\hat A_1 &= \hat{\kappa}_{i}^{a} + \hat{\kappa}_{\bar{i}}^{\bar{a}} \\
\hat A_2 &= \hat{\kappa}_{i\bar{i}}^{a\bar{a}} \\
\hat A_3 &= (\hat{n}_{a} - \hat{n}_{i} ) \hat{\kappa}_{\bar{i}}^{\bar{a}} + (\hat{n}_{\bar{a}} - \hat{n}_{\bar{i}} ) \hat{\kappa}_{i}^{a}  \\
\hat A_4 &= (\hat{n}_a - \hat{n}_i)^2 \hat{\kappa}_{\bar{i}}^{\bar{a}} + 
(\hat{n}_{\bar{a}}-\hat{n}_{\bar{i}})^2 \hat{\kappa}_{i}^{a} . 
\end{align}
It is easy to show that all these operators are singlet spherical tensor operators. This algebra 
can be decomposed to the center 
\bea
\hat A_C = [1-(\hn_a-\hn_i)^2]\kh_{\bar{i}}^{\bar{a}} + [1-(\hn_{\bar{a}}-\hn_{\bar{i}})^2]\kh_{i}^{a} 
\eea
 and the simple algebra isomorphic to $\LA{su}(2)$
\bea
\mathcal{A}_S = \{\hat A_2, \hat A_3/2, \hat A_4/2\}.
\eea
$\hat A_C$ zeroes any configuration in \eq{eq:FCI}, and thus one can use only elements of 
$\mathcal{A}_S$ to build the singlet wavefunction without the order dependence as
\begin{align}\label{eq:an}
\hat U(\boldsymbol{\tau})\ket{i\bar{i}} = \prod_{j=2}^4 e^{\tau_j \hat A_j} \ket{i\bar{i}} ,
\end{align}
where $\boldsymbol{\tau}=\{\tau_j\}$ are three optimization parameters 
(note that their optimal values can depend on the order of the exponents in the product).  
As a numerical test for this ansatz we optimized $\hat U(\boldsymbol{\tau})$ to reproduce
the wavefunction 
\bea
|\tilde{\Psi}\rangle =  \frac{1}{\sqrt{2}} \left( |a\bar{i}\rangle + |i\bar{a}\rangle \right),
\eea
which was shown to be unreachable using \eq{eq:P211}.\cite{Evangelista:kz} 
In addition, we tested all 6=3! different orderings of exponents in \eq{eq:an}
(e.g. $e^{\tau_4\hat A_4}e^{\tau_3\hat A_3}e^{\tau_2\hat A_2}$) and obtained 
 $|\langle \tilde{\Psi}|\hat U(\boldsymbol{\tau})\ket{i\bar{i}}| = 1$ for all of them. 

\section{Conclusions} \label{conclusion}

The order problem appears when one uses a partial product of non-commuting elements
of the $SU(2^N)$ Lie group to construct a unitary transformation in the VQE approach. Here, 
we have shown that elimination of the order dependence can be done by using Lie sub-groups
generated from corresponding Lie sub-algebras. Sub-algebras are obtained by considering all
commutators of some initial sub-set generated using heuristics for energy minimization. 
Sub-algebra construction increases the number of elements in the unitary transformation, 
to reduce the number of parameters for optimization without breaking the closure, 
we proposed reduction based on the system symmetry and analysis of the sub-algebra center. 
This reduction builds symmetry-adapted linear combinations 
of sub-algebra elements that commute with the symmetry operators.  

Some caution needs to be exercised in application of the proposed procedures. To eliminate the order problem 
completely, one needs to generate a sub-algebra containing all generators important for the energy lowering. 
This sub-algebra can be enormous so that its symmetry-adapted reduction is still quite large. 
The symmetry adaptation, even though reduces the number of parameters to optimize, constructs group 
elements that can require deep 
circuits to be implemented. This aspect requires further careful investigation.  
On the other hand, if one does not create large sub-algebras, imposing symmetrization can lead to empty sets of symmetrized 
generators (e.g., single fermionic excitation/de-excitation sets) and as a result removal of important operators. 

\section{Acknowledgements}
 A.F.I. is grateful to Lisa Jeffrey, Gustavo E. Scuseria, Francesco A. Evangelista, and Ilya G. Ryabinkin 
 for useful discussions and acknowledges financial support from the Google Quantum Research Program, 
 Early Researcher Award and the Natural Sciences and Engineering Research Council of Canada. 


\begin{thebibliography}{24}%
\makeatletter
\providecommand \@ifxundefined [1]{%
 \@ifx{#1\undefined}
}%
\providecommand \@ifnum [1]{%
 \ifnum #1\expandafter \@firstoftwo
 \else \expandafter \@secondoftwo
 \fi
}%
\providecommand \@ifx [1]{%
 \ifx #1\expandafter \@firstoftwo
 \else \expandafter \@secondoftwo
 \fi
}%
\providecommand \natexlab [1]{#1}%
\providecommand \enquote  [1]{``#1''}%
\providecommand \bibnamefont  [1]{#1}%
\providecommand \bibfnamefont [1]{#1}%
\providecommand \citenamefont [1]{#1}%
\providecommand \href@noop [0]{\@secondoftwo}%
\providecommand \href [0]{\begingroup \@sanitize@url \@href}%
\providecommand \@href[1]{\@@startlink{#1}\@@href}%
\providecommand \@@href[1]{\endgroup#1\@@endlink}%
\providecommand \@sanitize@url [0]{\catcode `\\12\catcode `\$12\catcode
  `\&12\catcode `\#12\catcode `\^12\catcode `\_12\catcode `\%12\relax}%
\providecommand \@@startlink[1]{}%
\providecommand \@@endlink[0]{}%
\providecommand \url  [0]{\begingroup\@sanitize@url \@url }%
\providecommand \@url [1]{\endgroup\@href {#1}{\urlprefix }}%
\providecommand \urlprefix  [0]{URL }%
\providecommand \Eprint [0]{\href }%
\providecommand \doibase [0]{http://dx.doi.org/}%
\providecommand \selectlanguage [0]{\@gobble}%
\providecommand \bibinfo  [0]{\@secondoftwo}%
\providecommand \bibfield  [0]{\@secondoftwo}%
\providecommand \translation [1]{[#1]}%
\providecommand \BibitemOpen [0]{}%
\providecommand \bibitemStop [0]{}%
\providecommand \bibitemNoStop [0]{.\EOS\space}%
\providecommand \EOS [0]{\spacefactor3000\relax}%
\providecommand \BibitemShut  [1]{\csname bibitem#1\endcsname}%
\let\auto@bib@innerbib\@empty
\bibitem [{\citenamefont {Peruzzo}\ \emph {et~al.}(2014)\citenamefont
  {Peruzzo}, \citenamefont {McClean}, \citenamefont {Shadbolt}, \citenamefont
  {Yung}, \citenamefont {Zhou}, \citenamefont {Love}, \citenamefont
  {Aspuru-Guzik},\ and\ \citenamefont {O'Brien}}]{Peruzzo2014}%
  \BibitemOpen
  \bibfield  {author} {\bibinfo {author} {\bibfnamefont {A.}~\bibnamefont
  {Peruzzo}}, \bibinfo {author} {\bibfnamefont {J.}~\bibnamefont {McClean}},
  \bibinfo {author} {\bibfnamefont {P.}~\bibnamefont {Shadbolt}}, \bibinfo
  {author} {\bibfnamefont {M.-H.}\ \bibnamefont {Yung}}, \bibinfo {author}
  {\bibfnamefont {X.-Q.}\ \bibnamefont {Zhou}}, \bibinfo {author}
  {\bibfnamefont {P.~J.}\ \bibnamefont {Love}}, \bibinfo {author}
  {\bibfnamefont {A.}~\bibnamefont {Aspuru-Guzik}}, \ and\ \bibinfo {author}
  {\bibfnamefont {J.~L.}\ \bibnamefont {O'Brien}},\ }\href@noop {} {\bibfield
  {journal} {\bibinfo  {journal} {Nat. Commun.}\ }\textbf {\bibinfo {volume}
  {5}},\ \bibinfo {pages} {4213} (\bibinfo {year} {2014})}\BibitemShut
  {NoStop}%
\bibitem [{\citenamefont {Gilmore}(2008{\natexlab{a}})}]{Gilmore_Exp}%
  \BibitemOpen
  \bibfield  {author} {\bibinfo {author} {\bibfnamefont {R.}~\bibnamefont
  {Gilmore}},\ }in\ \href@noop {} {\emph {\bibinfo {booktitle} {Lie Groups,
  Physics, and Geometry: An Introduction for Physicists, Engineers and
  Chemists}}}\ (\bibinfo  {publisher} {Cambridge University Press},\ \bibinfo
  {address} {Cambridge, UK},\ \bibinfo {year} {2008})\ p.~\bibinfo {pages}
  {57}\BibitemShut {NoStop}%
\bibitem [{\citenamefont {Knapp}(2002)}]{Knapp_expMsurj}%
  \BibitemOpen
  \bibfield  {author} {\bibinfo {author} {\bibfnamefont {A.~W.}\ \bibnamefont
  {Knapp}},\ }in\ \href@noop {} {\emph {\bibinfo {booktitle} {Lie groups beyond
  an introduction}}}\ (\bibinfo  {publisher} {Second Edition, Progress in
  Mathematics, Vol. 140},\ \bibinfo {address} {Birkhäuser, Boston},\ \bibinfo
  {year} {2002})\ p.\ \bibinfo {pages} {259}\BibitemShut {NoStop}%
\bibitem [{\citenamefont {Gilmore}(2008{\natexlab{b}})}]{Gilmore_BCH}%
  \BibitemOpen
  \bibfield  {author} {\bibinfo {author} {\bibfnamefont {R.}~\bibnamefont
  {Gilmore}},\ }in\ \href@noop {} {\emph {\bibinfo {booktitle} {Lie Groups,
  Physics, and Geometry: An Introduction for Physicists, Engineers and
  Chemists}}}\ (\bibinfo  {publisher} {Cambridge University Press},\ \bibinfo
  {address} {Cambridge, UK},\ \bibinfo {year} {2008})\ p.\ \bibinfo {pages}
  {108}\BibitemShut {NoStop}%
\bibitem [{Note1()}]{Note1}%
  \BibitemOpen
  \bibinfo {note} {Note that quantum computers cannot perform $\exp (i\tau _k
  \protect \mathaccentV {hat}05EP_k)$ as elementary circuit operations (gates)
  but there are compilers that present such operations as sequences of
  universal gates containing not more than polynomial in $N$ number of gates.
  Thus, we assume that polynomial in the number of $P_k$'s algorithms is a
  desirable goal.}\BibitemShut {Stop}%
\bibitem [{\citenamefont {Lee}\ \emph {et~al.}(2019)\citenamefont {Lee},
  \citenamefont {Huggins}, \citenamefont {Head-Gordon},\ and\ \citenamefont
  {Whaley}}]{Lee:2019/jctc/311}%
  \BibitemOpen
  \bibfield  {author} {\bibinfo {author} {\bibfnamefont {J.}~\bibnamefont
  {Lee}}, \bibinfo {author} {\bibfnamefont {W.~J.}\ \bibnamefont {Huggins}},
  \bibinfo {author} {\bibfnamefont {M.}~\bibnamefont {Head-Gordon}}, \ and\
  \bibinfo {author} {\bibfnamefont {K.~B.}\ \bibnamefont {Whaley}},\ }\href
  {\doibase 10.1021/acs.jctc.8b01004} {\bibfield  {journal} {\bibinfo
  {journal} {J. Chem. Theory Comput.}\ }\textbf {\bibinfo {volume} {15}},\
  \bibinfo {pages} {311} (\bibinfo {year} {2019})}\BibitemShut {NoStop}%
\bibitem [{\citenamefont {Sokolov}\ \emph {et~al.}(2019)\citenamefont
  {Sokolov}, \citenamefont {Barkoutsos}, \citenamefont {Ollitrault},
  \citenamefont {Greenberg}, \citenamefont {Rice}, \citenamefont {Pistoia},\
  and\ \citenamefont {Tavernelli}}]{Sokolov2019}%
  \BibitemOpen
  \bibfield  {author} {\bibinfo {author} {\bibfnamefont {I.}~\bibnamefont
  {Sokolov}}, \bibinfo {author} {\bibfnamefont {P.~K.}\ \bibnamefont
  {Barkoutsos}}, \bibinfo {author} {\bibfnamefont {P.~J.}\ \bibnamefont
  {Ollitrault}}, \bibinfo {author} {\bibfnamefont {D.}~\bibnamefont
  {Greenberg}}, \bibinfo {author} {\bibfnamefont {J.}~\bibnamefont {Rice}},
  \bibinfo {author} {\bibfnamefont {M.}~\bibnamefont {Pistoia}}, \ and\
  \bibinfo {author} {\bibfnamefont {I.}~\bibnamefont {Tavernelli}},\
  }\href@noop {} {\bibfield  {journal} {\bibinfo  {journal} {arXiv.org}\ }
  (\bibinfo {year} {2019})},\ \Eprint {http://arxiv.org/abs/1911.10864}
  {arXiv:1911.10864 [quant-ph]} \BibitemShut {NoStop}%
\bibitem [{\citenamefont {Evangelista}\ \emph {et~al.}(2019)\citenamefont
  {Evangelista}, \citenamefont {Chan},\ and\ \citenamefont
  {Scuseria}}]{Evangelista:kz}%
  \BibitemOpen
  \bibfield  {author} {\bibinfo {author} {\bibfnamefont {F.~A.}\ \bibnamefont
  {Evangelista}}, \bibinfo {author} {\bibfnamefont {G.~K.-L.}\ \bibnamefont
  {Chan}}, \ and\ \bibinfo {author} {\bibfnamefont {G.~E.}\ \bibnamefont
  {Scuseria}},\ }\href {\doibase 10.1063/1.5133059} {\bibfield  {journal}
  {\bibinfo  {journal} {J. Chem. Phys.}\ }\textbf {\bibinfo {volume} {151}},\
  \bibinfo {pages} {244112} (\bibinfo {year} {2019})}\BibitemShut {NoStop}%
\bibitem [{\citenamefont {{Mizukami}}\ \emph {et~al.}(2019)\citenamefont
  {{Mizukami}}, \citenamefont {{Mitarai}}, \citenamefont {{Nakagawa}},
  \citenamefont {{Yamamoto}}, \citenamefont {{Yan}},\ and\ \citenamefont
  {{Ohnishi}}}]{Mizukami2019}%
  \BibitemOpen
  \bibfield  {author} {\bibinfo {author} {\bibfnamefont {W.}~\bibnamefont
  {{Mizukami}}}, \bibinfo {author} {\bibfnamefont {K.}~\bibnamefont
  {{Mitarai}}}, \bibinfo {author} {\bibfnamefont {Y.~O.}\ \bibnamefont
  {{Nakagawa}}}, \bibinfo {author} {\bibfnamefont {T.}~\bibnamefont
  {{Yamamoto}}}, \bibinfo {author} {\bibfnamefont {T.}~\bibnamefont {{Yan}}}, \
  and\ \bibinfo {author} {\bibfnamefont {Y.-y.}\ \bibnamefont {{Ohnishi}}},\
  }\href@noop {} {\bibfield  {journal} {\bibinfo  {journal} {arXiv e-prints}\ }
  (\bibinfo {year} {2019})},\ \Eprint {http://arxiv.org/abs/1910.11526}
  {arXiv:1910.11526 [cond-mat.str-el]} \BibitemShut {NoStop}%
\bibitem [{\citenamefont {Hempel}\ \emph {et~al.}(2018)\citenamefont {Hempel},
  \citenamefont {Maier}, \citenamefont {Romero}, \citenamefont {McClean},
  \citenamefont {Monz}, \citenamefont {Shen}, \citenamefont {Jurcevic},
  \citenamefont {Lanyon}, \citenamefont {Love}, \citenamefont {Babbush},
  \citenamefont {Aspuru-Guzik}, \citenamefont {Blatt},\ and\ \citenamefont
  {Roos}}]{Hempel2018}%
  \BibitemOpen
  \bibfield  {author} {\bibinfo {author} {\bibfnamefont {C.}~\bibnamefont
  {Hempel}}, \bibinfo {author} {\bibfnamefont {C.}~\bibnamefont {Maier}},
  \bibinfo {author} {\bibfnamefont {J.}~\bibnamefont {Romero}}, \bibinfo
  {author} {\bibfnamefont {J.}~\bibnamefont {McClean}}, \bibinfo {author}
  {\bibfnamefont {T.}~\bibnamefont {Monz}}, \bibinfo {author} {\bibfnamefont
  {H.}~\bibnamefont {Shen}}, \bibinfo {author} {\bibfnamefont {P.}~\bibnamefont
  {Jurcevic}}, \bibinfo {author} {\bibfnamefont {B.~P.}\ \bibnamefont
  {Lanyon}}, \bibinfo {author} {\bibfnamefont {P.}~\bibnamefont {Love}},
  \bibinfo {author} {\bibfnamefont {R.}~\bibnamefont {Babbush}}, \bibinfo
  {author} {\bibfnamefont {A.}~\bibnamefont {Aspuru-Guzik}}, \bibinfo {author}
  {\bibfnamefont {R.}~\bibnamefont {Blatt}}, \ and\ \bibinfo {author}
  {\bibfnamefont {C.~F.}\ \bibnamefont {Roos}},\ }\href@noop {} {\bibfield
  {journal} {\bibinfo  {journal} {Phys. Rev. X}\ }\textbf {\bibinfo {volume}
  {8}},\ \bibinfo {pages} {31022} (\bibinfo {year} {2018})}\BibitemShut
  {NoStop}%
\bibitem [{\citenamefont {Ryabinkin}\ \emph {et~al.}(2018)\citenamefont
  {Ryabinkin}, \citenamefont {Yen}, \citenamefont {Genin},\ and\ \citenamefont
  {Izmaylov}}]{Ryabinkin2018}%
  \BibitemOpen
  \bibfield  {author} {\bibinfo {author} {\bibfnamefont {I.~G.}\ \bibnamefont
  {Ryabinkin}}, \bibinfo {author} {\bibfnamefont {T.-C.}\ \bibnamefont {Yen}},
  \bibinfo {author} {\bibfnamefont {S.~N.}\ \bibnamefont {Genin}}, \ and\
  \bibinfo {author} {\bibfnamefont {A.~F.}\ \bibnamefont {Izmaylov}},\
  }\href@noop {} {\bibfield  {journal} {\bibinfo  {journal} {J. Chem. Theory
  Comput.}\ }\textbf {\bibinfo {volume} {14}},\ \bibinfo {pages} {6317}
  (\bibinfo {year} {2018})}\BibitemShut {NoStop}%
\bibitem [{\citenamefont {Grimsley}\ \emph
  {et~al.}(2019{\natexlab{a}})\citenamefont {Grimsley}, \citenamefont
  {Economou}, \citenamefont {Barnes},\ and\ \citenamefont
  {Mayhall}}]{Grimsley2019}%
  \BibitemOpen
  \bibfield  {author} {\bibinfo {author} {\bibfnamefont {H.~R.}\ \bibnamefont
  {Grimsley}}, \bibinfo {author} {\bibfnamefont {S.~E.}\ \bibnamefont
  {Economou}}, \bibinfo {author} {\bibfnamefont {E.}~\bibnamefont {Barnes}}, \
  and\ \bibinfo {author} {\bibfnamefont {N.~J.}\ \bibnamefont {Mayhall}},\
  }\href@noop {} {\bibfield  {journal} {\bibinfo  {journal} {Nat. Commun.}\
  }\textbf {\bibinfo {volume} {10}},\ \bibinfo {pages} {3007} (\bibinfo {year}
  {2019}{\natexlab{a}})}\BibitemShut {NoStop}%
\bibitem [{\citenamefont {Tang}\ \emph {et~al.}(2019)\citenamefont {Tang},
  \citenamefont {Barnes}, \citenamefont {Grimsley}, \citenamefont {Mayhall},\
  and\ \citenamefont {Economou}}]{qAdapt}%
  \BibitemOpen
  \bibfield  {author} {\bibinfo {author} {\bibfnamefont {H.~L.}\ \bibnamefont
  {Tang}}, \bibinfo {author} {\bibfnamefont {E.}~\bibnamefont {Barnes}},
  \bibinfo {author} {\bibfnamefont {H.~R.}\ \bibnamefont {Grimsley}}, \bibinfo
  {author} {\bibfnamefont {N.~J.}\ \bibnamefont {Mayhall}}, \ and\ \bibinfo
  {author} {\bibfnamefont {S.~E.}\ \bibnamefont {Economou}},\ }\href@noop {}
  {\bibfield  {journal} {\bibinfo  {journal} {arXiv.org}\ } (\bibinfo {year}
  {2019})},\ \Eprint {http://arxiv.org/abs/1911.10205v1} {1911.10205v1}
  \BibitemShut {NoStop}%
\bibitem [{\citenamefont {Grimsley}\ \emph
  {et~al.}(2019{\natexlab{b}})\citenamefont {Grimsley}, \citenamefont
  {Claudino}, \citenamefont {Economou}, \citenamefont {Barnes},\ and\
  \citenamefont {Mayhall}}]{Grimsley:2019wz}%
  \BibitemOpen
  \bibfield  {author} {\bibinfo {author} {\bibfnamefont {H.~R.}\ \bibnamefont
  {Grimsley}}, \bibinfo {author} {\bibfnamefont {D.}~\bibnamefont {Claudino}},
  \bibinfo {author} {\bibfnamefont {S.~E.}\ \bibnamefont {Economou}}, \bibinfo
  {author} {\bibfnamefont {E.}~\bibnamefont {Barnes}}, \ and\ \bibinfo {author}
  {\bibfnamefont {N.~J.}\ \bibnamefont {Mayhall}},\ }\href@noop {} {\bibfield
  {journal} {\bibinfo  {journal} {arXiv.org}\ } (\bibinfo {year}
  {2019}{\natexlab{b}})},\ \Eprint {http://arxiv.org/abs/1910.10329v1}
  {1910.10329v1} \BibitemShut {NoStop}%
\bibitem [{\citenamefont {Tranter}\ \emph {et~al.}(2019)\citenamefont
  {Tranter}, \citenamefont {Love}, \citenamefont {Mintert}, \citenamefont
  {Wiebe},\ and\ \citenamefont {Coveney}}]{Tranter:2019im}%
  \BibitemOpen
  \bibfield  {author} {\bibinfo {author} {\bibfnamefont {A.}~\bibnamefont
  {Tranter}}, \bibinfo {author} {\bibfnamefont {P.~J.}\ \bibnamefont {Love}},
  \bibinfo {author} {\bibfnamefont {F.}~\bibnamefont {Mintert}}, \bibinfo
  {author} {\bibfnamefont {N.}~\bibnamefont {Wiebe}}, \ and\ \bibinfo {author}
  {\bibfnamefont {P.~V.}\ \bibnamefont {Coveney}},\ }\href@noop {} {\bibfield
  {journal} {\bibinfo  {journal} {Entropy}\ }\textbf {\bibinfo {volume} {21}},\
  \bibinfo {pages} {1218} (\bibinfo {year} {2019})}\BibitemShut {NoStop}%
\bibitem [{\citenamefont {Gilmore}(2008{\natexlab{c}})}]{Gilmore:2008}%
  \BibitemOpen
  \bibfield  {author} {\bibinfo {author} {\bibfnamefont {R.}~\bibnamefont
  {Gilmore}},\ }\href@noop {} {\emph {\bibinfo {title} {{Lie Groups, Physics,
  and Geometry: An Introduction for Physicists, Engineers and Chemists}}}}\
  (\bibinfo  {publisher} {Cambridge University Press},\ \bibinfo {year}
  {2008})\BibitemShut {NoStop}%
\bibitem [{\citenamefont {Barut}\ and\ \citenamefont
  {Raczka}(1980)}]{Barut_p17}%
  \BibitemOpen
  \bibfield  {author} {\bibinfo {author} {\bibfnamefont {A.~O.}\ \bibnamefont
  {Barut}}\ and\ \bibinfo {author} {\bibfnamefont {R.}~\bibnamefont {Raczka}},\
  }in\ \href@noop {} {\emph {\bibinfo {booktitle} {{Theory of Group
  Representations and Applications}}}}\ (\bibinfo  {publisher} {Polish
  Scientific Publisher},\ \bibinfo {address} {Warszawa, Poland},\ \bibinfo
  {year} {1980})\ p.~\bibinfo {pages} {17}\BibitemShut {NoStop}%
\bibitem [{\citenamefont {Romero}\ \emph {et~al.}(2018)\citenamefont {Romero},
  \citenamefont {Babbush}, \citenamefont {McClean}, \citenamefont {Hempel},
  \citenamefont {Love},\ and\ \citenamefont {Aspuru-Guzik}}]{Romero2018}%
  \BibitemOpen
  \bibfield  {author} {\bibinfo {author} {\bibfnamefont {J.}~\bibnamefont
  {Romero}}, \bibinfo {author} {\bibfnamefont {R.}~\bibnamefont {Babbush}},
  \bibinfo {author} {\bibfnamefont {J.~R.}\ \bibnamefont {McClean}}, \bibinfo
  {author} {\bibfnamefont {C.}~\bibnamefont {Hempel}}, \bibinfo {author}
  {\bibfnamefont {P.~J.}\ \bibnamefont {Love}}, \ and\ \bibinfo {author}
  {\bibfnamefont {A.}~\bibnamefont {Aspuru-Guzik}},\ }\href@noop {} {\bibfield
  {journal} {\bibinfo  {journal} {Quantum Sci. and Technol.}\ }\textbf
  {\bibinfo {volume} {4}},\ \bibinfo {pages} {014008} (\bibinfo {year}
  {2018})}\BibitemShut {NoStop}%
\bibitem [{\citenamefont {Helgaker}\ \emph {et~al.}(2000)\citenamefont
  {Helgaker}, \citenamefont {J{\o}rgensen},\ and\ \citenamefont
  {Olsen}}]{Helgaker}%
  \BibitemOpen
  \bibfield  {author} {\bibinfo {author} {\bibfnamefont {T.}~\bibnamefont
  {Helgaker}}, \bibinfo {author} {\bibfnamefont {P.}~\bibnamefont
  {J{\o}rgensen}}, \ and\ \bibinfo {author} {\bibfnamefont {J.}~\bibnamefont
  {Olsen}},\ }\href@noop {} {\emph {\bibinfo {title} {{Molecular
  Electronic-Structure Theory}}}}\ (\bibinfo  {publisher} {John Wiley {\&}
  Sons, Ltd},\ \bibinfo {address} {Chichester, UK},\ \bibinfo {year}
  {2000})\BibitemShut {NoStop}%
\bibitem [{\citenamefont {Li}\ and\ \citenamefont {Paldus}(2014)}]{Paldus}%
  \BibitemOpen
  \bibfield  {author} {\bibinfo {author} {\bibfnamefont {X.}~\bibnamefont
  {Li}}\ and\ \bibinfo {author} {\bibfnamefont {J.}~\bibnamefont {Paldus}},\
  }\href@noop {} {\bibfield  {journal} {\bibinfo  {journal} {Theor. Chem.
  Acc.}\ }\textbf {\bibinfo {volume} {133}},\ \bibinfo {pages} {1467} (\bibinfo
  {year} {2014})}\BibitemShut {NoStop}%
\bibitem [{\citenamefont {Shavitt}(2009)}]{Shavitt}%
  \BibitemOpen
  \bibfield  {author} {\bibinfo {author} {\bibfnamefont {I.}~\bibnamefont
  {Shavitt}},\ }\href {\doibase 10.1002/qua.560140803} {\bibfield  {journal}
  {\bibinfo  {journal} {Int. J. Quant. Chem.}\ }\textbf {\bibinfo {volume}
  {14}},\ \bibinfo {pages} {5 } (\bibinfo {year} {2009})}\BibitemShut {NoStop}%
\bibitem [{\citenamefont {Ryabinkin}\ \emph {et~al.}(2020)\citenamefont
  {Ryabinkin}, \citenamefont {Lang}, \citenamefont {Genin},\ and\ \citenamefont
  {Izmaylov}}]{Ryabinkin2019b}%
  \BibitemOpen
  \bibfield  {author} {\bibinfo {author} {\bibfnamefont {I.~G.}\ \bibnamefont
  {Ryabinkin}}, \bibinfo {author} {\bibfnamefont {R.~A.}\ \bibnamefont {Lang}},
  \bibinfo {author} {\bibfnamefont {S.~N.}\ \bibnamefont {Genin}}, \ and\
  \bibinfo {author} {\bibfnamefont {A.~F.}\ \bibnamefont {Izmaylov}},\ }\href
  {\doibase 10.1021/acs.jctc.9b01084} {\bibfield  {journal} {\bibinfo
  {journal} {J. Chem. Theory Comput.}\ }\textbf {\bibinfo {volume} {16}},\
  \bibinfo {pages} {1055} (\bibinfo {year} {2020})}\BibitemShut {NoStop}%
\bibitem [{\citenamefont {{Zhao}}\ \emph {et~al.}(2019)\citenamefont {{Zhao}},
  \citenamefont {{Tranter}}, \citenamefont {{Kirby}}, \citenamefont {{Ung}},
  \citenamefont {{Miyake}},\ and\ \citenamefont {{Love}}}]{Zhao2019}%
  \BibitemOpen
  \bibfield  {author} {\bibinfo {author} {\bibfnamefont {A.}~\bibnamefont
  {{Zhao}}}, \bibinfo {author} {\bibfnamefont {A.}~\bibnamefont {{Tranter}}},
  \bibinfo {author} {\bibfnamefont {W.~M.}\ \bibnamefont {{Kirby}}}, \bibinfo
  {author} {\bibfnamefont {S.~F.}\ \bibnamefont {{Ung}}}, \bibinfo {author}
  {\bibfnamefont {A.}~\bibnamefont {{Miyake}}}, \ and\ \bibinfo {author}
  {\bibfnamefont {P.}~\bibnamefont {{Love}}},\ }\href@noop {} {\bibfield
  {journal} {\bibinfo  {journal} {arXiv e-prints}\ } (\bibinfo {year}
  {2019})},\ \Eprint {http://arxiv.org/abs/1908.08067} {arXiv:1908.08067
  [quant-ph]} \BibitemShut {NoStop}%
\bibitem [{\citenamefont {Paldus}\ and\ \citenamefont
  {Sarma}(1985)}]{PaldusSarma1985}%
  \BibitemOpen
  \bibfield  {author} {\bibinfo {author} {\bibfnamefont {J.}~\bibnamefont
  {Paldus}}\ and\ \bibinfo {author} {\bibfnamefont {C.~R.}\ \bibnamefont
  {Sarma}},\ }\href@noop {} {\bibfield  {journal} {\bibinfo  {journal} {J.
  Chem. Phys.}\ }\textbf {\bibinfo {volume} {83}},\ \bibinfo {pages} {5135}
  (\bibinfo {year} {1985})}\BibitemShut {NoStop}%
\end{thebibliography}

%

\end{document}